# Longitudinal Impact of Tobacco Use and Social Determinants on Respiratory Health Disparities Among Louisiana Medicaid Enrollees

*Yead Rahman, MS[1,] Prerna Dua, PhD[2]

[1] Department of Electrical Engineering, College of Engineering and Science, Louisiana Tech University, Ruston, Louisiana, United States.

[2] Department of Biomedical and Health Informatics, School of Medicine, University of Missouri, Kansas City, Missouri, United States.

**Corresponding author**:

Yead Rahman, MS

Electrical Engineering, Louisiana Tech University

Ruston, LA, United States

Email: yra006@latech.edu

**Abstract**

Tobacco use remains a leading preventable contributor to serious health conditions in the United States, notably chronic obstructive pulmonary disease (COPD) and severe COVID-19 complications. Within Louisiana's Medicaid population, tobacco use prevalence is particularly high compared to privately insured groups, yet its full impact on long-term outcomes is not fully understood.

This study aimed to investigate how tobacco use, in conjunction with demographic and clinical risk factors, influences the incidence of COPD and COVID-19 among Medicaid enrollees over time.

We analyzed Louisiana Department of Health data from January 2020 to February 2023. Chi-square tests were conducted to provide descriptive statistics, and multivariate logistic regression models were applied across three discrete waves to assess both cross-sectional and longitudinal associations between risk factors and disease outcomes. Enrollees without baseline diagnoses of COPD or COVID-19 were followed to determine new-onset cases in subsequent waves. Adjusted odds ratios (AOR) were calculated after controlling for socio-demographic variables, comorbidities, and healthcare utilization patterns.

Tobacco use emerged as a significant independent predictor of both COPD (AOR = 1.12, 95% CI = 1.05–1.18) and COVID-19 (AOR = 1.66, 95% CI = 1.15–2.38). Additional risk factors-such as older age, gender, region, and pre-existing health conditions-also showed significant associations with higher incidence rates of COPD and COVID-19.

By linking tobacco use, demographic disparities, and comorbidities to an increased risk of COPD and COVID-19, this study underscores the urgent need for targeted tobacco cessation efforts and prevention strategies within this underserved population.

**Keywords**: Tobacco use, social determinants of health, Medicaid enrollees, COPD, COVID-19, statistical analysis

## 1. Introduction

Tobacco use, encompassing both combustible and non-combustible forms, remains a primary preventable contributor to chronic obstructive pulmonary disease (COPD), heightened COVID-19 severity, and other chronic health outcomes in the United States. COPD imposes a significant health burden, ranking as a leading cause of death nationally [1]. Recent estimates point to 14.2 million diagnosed cases [2], with 138,825 related deaths in 2021 alone-placing it sixth in U.S. mortality statistics [3]. Meanwhile, COVID-19 was declared a global pandemic by the World Health Organization (WHO) [4], and has resulted in approximately 1.2 million fatalities across the country as of April 20, 2024 [5].

Evidence indicates that tobacco smoke contains numerous harmful substances, including nitrosamines and polycyclic aromatic hydrocarbons, known as well-established carcinogens [6]. One mechanism for increased COVID-19 susceptibility among smokers involves the upregulation of the ACE2 receptor [7]. Moreover, cigarette smoke carries more than 4,000 toxic agents, many of which damage lung tissue and blood vessels, driving COPD progression [8,9]. E-cigarette aerosols also contain acrolein, a byproduct of e-liquid combustion, which can weaken immune defenses and exacerbate lung infections [10]. Beyond these biological pathways, social factors such as rural residence create additional healthcare disparities for COPD patients, especially among marginalized populations [11].

Louisiana notably records a tobacco use prevalence of 18.9% [12]. Nationally, Medicaid beneficiaries exhibit higher rates of cigarette smoking compared to individuals with private insurance (23.9% vs. 10.5%) [13], coupled with increased incidences of chronic disease, frequent clinical visits, and significant psychological distress with over half of smokers report severe mental health challenges [14]. Additionally, an estimated 11% of Medicaid spending (95% CI: 0.4%–17.0%) is tied to smoking [15].

Despite the extensive studies of chronic disease burden among Medicaid populations [16], the interplay between behavioral risk, alongside social determinants and long-term health outcomes remains underexplored. Tobacco use has been broadly acknowledged as a major determinant of adverse health trajectories, yet its interaction with socioeconomic and clinical variables requires further elucidation. Existing evidence underscores the importance of chronic conditions and demographic factors in driving COVID-19 mortality, highlighting the need to assess how tobacco consumption may worsen respiratory disorders like COPD and compound disparities among Medicaid enrollees [17]. Socioeconomic determinants significantly shape life expectancy and general health outcomes, aligning with this study's goal of examining the wider implications of tobacco use on chronic disease progression within vulnerable populations [18]. Additionally, prior work demonstrates the utility of Medicaid claims data in identifying environmental and demographic drivers of disease, reinforcing our approach to tracking trends in tobacco-related morbidity [19]. Targeted intervention for high-risk subgroups is vital, informing our focus on Medicaid beneficiaries prone to COPD and COVID-19 complications [20]. Finally, robust statistical methodologies remain essential in health outcomes research, supporting our choice of advanced modeling to evaluate the long-term impact of tobacco use across Medicaid-enrolled groups [21]. Integrating these diverse insights, this study presents a data-driven framework for informing public health strategies and optimizing healthcare resource allocation to assist at-risk Medicaid beneficiaries. Using data from 2020 to 2022, the following analysis specifically investigates the longitudinal impact of tobacco use, clinical features, and sociodemographic determinants on adverse health outcomes, including COPD and COVID-19, in the Louisiana Medicaid population.

## 2. Materials and Methods

### 2.1 Study Population

This research leveraged a large-scale claims dataset representing 168,542 unique individuals enrolled in Louisiana Medicaid from January 2020 to February 2023. By focusing on enrollees aged 15 to 75 years at the start of the measurement year who had at least one recorded diagnosis, we established a robust analytic base for tracking longitudinal health outcomes. To capture changes over time, we subdivided the observation period into three distinct waves: Wave-1 (January 1–December 31, 2020), Wave-2 (January 1–December 31, 2021), and Wave-3 (January 1, 2022–February 28, 2023). Figure 1(A) depicts the distribution of participants across these intervals, supporting our wave-based modeling approach and enabling trend analysis of key health metrics.

### 2.1.1 Ethical Considerations

The Louisiana Department of Health provided all data in an anonymized format, removing direct and indirect identifiers to protect participant privacy. Because the final dataset contained no personally identifiable information, this study was deemed exempt from Institutional Review Board (IRB) review. By ensuring data confidentiality, our analytical framework aligns with ethical standards for secondary health data research.

## 2.2 Measures

All clinical diagnoses, including tobacco use status, COVID-19 infection, COPD, and comorbidities such as coronary artery disease (CAD) and Hypertensive Disease, were determined through International Classification of Diseases, Tenth Revision (ICD-10) codes [22]. This code-driven identification strategy ensured uniform categorization of health conditions across all three waves and allowed for consistent tracking of newly emerging or persistent diagnoses.

In line with our longitudinal analytic design, any enrollee with an ICD-10 code related to tobacco use recorded during Wave-1 was classified as a new ("incident") tobacco user, encompassing both combustible and non-combustible products. Parallel procedures applied to health conditions such as COVID-19 and COPD- patients were coded as positive for a disease if at least one relevant ICD-10 code appeared during a given wave. Baseline demographic data collected at Wave-1 included age (grouped in 10-year intervals between 16 and 75), self-reported gender (male or female), race (White, African American, or other), and residential region (New Orleans, Baton Rouge, Thibodeaux, Lafayette, Lake Charles, Alexandria, Shreveport, Monroe, or Mandeville). From a healthcare analytics perspective, this code-based framework enabled us to systematically integrate individual- and population-level risk factors, facilitating robust statistical modeling and trend detection in subsequent analysis.

## 2.3 Longitudinal Modeling

To investigate how tobacco use influences the onset of COPD and COVID-19 over time, we developed two separate longitudinal models- one targeting COPD incidence and the other

COVID-19 incidence. Figure 1(B) provides a visual overview of this modeling framework, illustrating the variables and analytic steps.

**2.3.1 Data Preprocessing**

This study leveraged an extensive dataset of approximately 6.3 million records sourced from the Louisiana Department of Health, covering January 1, 2020, through February 28, 2023. The data were stratified into three distinct time periods: Wave-1 (January-December 2020), Wave-2 (January–December 2021), and Wave-3 (January 2022-February 2023). Consistent feature engineering procedures were employed in each wave to classify key diagnoses- tobacco use, COVID-19, COPD, coronary artery disease (CAD), and hypertensive disease (HTN)- according to ICD-10 codes.

Any patient identified with COPD or COVID-19 at Wave-1 was excluded from subsequent analyses of that specific outcome to allow a clear assessment of incident disease. For each participant, the number of clinical services received in each wave was also compiled, reflecting total healthcare utilization. Demographic indicators such as age group (16-75 in six categories), gender (male/female), race (White/African American/other), regions, and disability status were converted to binary or dummy variables depending on the number of categories they have. After excluding a small number of observations with missing values in critical fields, outliers in numerical variables were removed using the interquartile range (IQR) approach to ensure data integrity.

Following data cleaning, patients without COPD or COVID-19 at baseline were merged across Waves 2 and 3 using unique identifiers. We then created a binary indicator denoting the emergence of COPD or COVID-19 in either Wave 2 or Wave 3. This consolidated outcome facilitated a more robust longitudinal analysis by capturing incident cases throughout the follow-up period, rather than focusing on a single wave.

**2.3.2 Model Construction**

We employed logistic regression to model the presence or absence of COPD and COVID-19 in Waves 2 or 3, given the binary nature of these outcomes [23]. The study utilized the logistic regression function from Python's statistical model's library.[24] This approach estimated the probability of disease onset based on predictor variables encompassing tobacco use, comorbidities (CAD, HTN), healthcare utilization (number of clinical services), and a full range of demographic and regional controls. By including all relevant covariates, we reduced the likelihood of omitted-variable bias and more accurately isolated the independent effect of tobacco use on subsequent disease [25].

**2.3.3 Model Training and Evaluation**

Using Python's statistical model, each logistic regression model was trained to assess how well the predictor set explained new cases of COPD or COVID-19 over time. Model performance was quantified with receiver operating characteristic (ROC) curves, and the corresponding area under the ROC curve (AUROC), as shown in Figure 1(c), served as our primary metric for predictive accuracy. We further examined the resulting coefficients, standard errors, p-values, and confidence intervals to determine which explanatory variables significantly contributed to disease risk. This combination of performance metrics and inferential tests helped validate the robustness of our wave-based analytic strategy and provided actionable insights into the factors driving COPD and COVID-19 incidence within the Louisiana Medicaid population.

**2.3.4 Statistical Analysis**

We began by examining the distributions of COPD and COVID-19 across two-time frames: at Wave-1 baseline and in the combined Wave-2 and Wave-3 follow-up. Chi-square tests were employed to compare group differences in these dependent variables, providing an initial snapshot of disease prevalence and potential associations with demographic or clinical factors. To evaluate how comorbidities (e.g., hypertension, coronary artery disease) independently affected COPD or COVID-19 presence at baseline (Wave-1), we ran cross-sectional logistic regressions. By controlling for additional demographic and clinical covariates, this approach isolated the effect of key comorbid conditions on initial disease risk.

Recognizing the need for a more dynamic view of disease progression, we developed two separate longitudinal logistic regression models—one focused on COPD incidence and the other on COVID-19 incidence over a two-year follow-up. Patients who did not present with COPD or COVID-19 at baseline were tracked during Waves 2 and 3, effectively capturing new diagnoses over time. Waves 2 and 3 were combined in each model to enhance statistical power and increase the number of events observed.

**COPD Incidence**: For individuals free of COPD at Wave-1, we tested whether tobacco use at baseline predicted new-onset COPD in Waves 2 or 3. Covariates included a prior COVID-19 diagnosis, coronary artery disease, hypertension, clinical service utilization (total visits), and demographic factors (age group, region, gender, race).

**COVID-19 Incidence:** For those uninfected at Wave-1, a parallel logistic regression estimated how baseline tobacco use influenced incident COVID-19 in Waves 2 or 3, controlling for the same clinical and demographic variables.

Binary indicators for tobacco use, disability status, COPD, COVID-19, coronary artery disease, hypertension, gender, and regional groupings were created. Age (in six categories spanning 15–75 years) and race (White, African American, or other) were treated as categorical variables, each transformed into dummy codes for regression. Reference categories consisted of the youngest age group (15–25), White race, non-use of tobacco, and the New Orleans region.

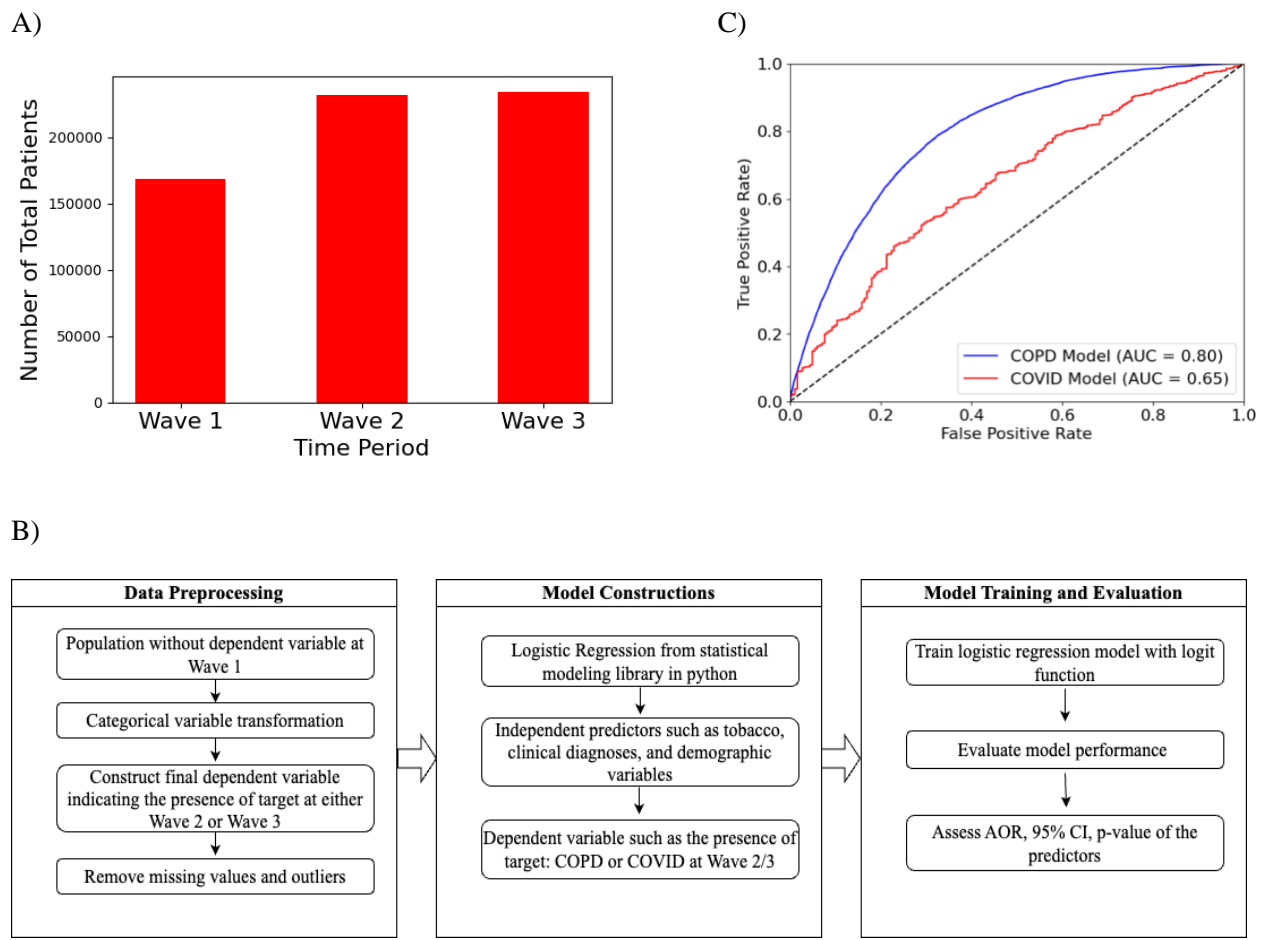

**Figure 1**: Outline of the Procedure: (A) Study populations were divided into three waves based on clinical service dates. (B) Two longitudinal models were created using Logistic Regression from Python's statistical model library. (C) Model performance was evaluated using receiver operating characteristic (ROC) curves.

Consequently, the logistic coefficients represent the change in the log-odds of disease when comparing each category to its respective reference group. This comprehensive specification allowed us to identify independent risk factors driving COPD and COVID-19 outcomes in a Medicaid population, thereby aiding targeted public health interventions and policy development.

## 3. Results Analysis

Table 1 summarizes key clinical and demographic attributes of the 168,542 Medicaid enrollees at Wave-1 (January 1–December 31, 2020). Notably, 21,708 individuals (12.88%) had COPD and 19,871 (11.79%) had COVID-19, underscoring the sizable group free of these conditions at baseline—and therefore eligible for subsequent longitudinal analyses. In addition, 76,730 enrollees (45.53%) were identified as tobacco users, suggesting that nearly half of the study population smoked or used other tobacco products.

Chi-square tests stratified by COPD and COVID-19 status at baseline (Wave-1) and at combined Waves 2 and 3 reveal several noteworthy patterns. Both Table 2.1 and Table 2.2 indicate that none of the individuals with a tobacco-use diagnosis at baseline also had COPD or COVID-19 at that time; however, among those free of COPD at baseline, 3,376 (p<0.001) went on to develop it in Waves 2 or 3 as shown in Table 2.1. Although the chi-square test did not detect a significant association between tobacco use at baseline and COVID-19 over follow-up as shown in Table 2.2, subsequent logistic regression models showed otherwise.

Table 3 presents cross-sectional logistic regression findings for Wave-1, examining associations between demographic/clinical factors and both COPD (left columns) and COVID-19 (right columns). For COPD, hypertension (AOR=1.75, 95% CI=1.67–1.85), CAD (AOR=1.18, 95% CI=1.12–1.25), and higher clinical service utilization (AOR=1.05, 95% CI=1.04–1.05) emerged as significant risk factors. Being in the 46–55 age group (AOR=4.02, 95% CI=3.00–5.38) and residing in the Lake Charles region (AOR=1.66, 95% CI=1.55–1.78) further elevated COPD risk.

Table 1 Demographic, Clinical, and Tobacco Use Variables at wave-1 (2020-01-01 to 2020-12-31) Baseline (n=168542)

| Variables | Weighted % |
|---|---|
| **Tobacco Use** | |
| Yes | 45.53 |
| No | 54.47 |
| **COVID-19 Disease** | |
| Yes | 11.79 |
| No | 88.21 |
| **Diabetes Disease** | |
| Yes | 41.37 |
| No | 58.63 |
| **COPD Disease** | |
| Yes | 12.88 |
| No | 87.12 |
| **CAD Disease** | |
| Yes | 5.51 |
| No | 94.49 |
| **Hypertension** | |
| Yes | 6.25 |
| No | 93.75 |
| **Demographic** | |
| **Age (years)** | |
| 16-25 | 14.24 |
| 26-35 | 23.16 |
| 36-45 | 20.70 |
| 46-55 | 18.82 |
| 56-65 | 19.04 |
| 66-75 | 4.04 |
| **Gender** | |
| Male | 39.79 |
| Female | 60.21 |
| **Race** | |
| White | 48.54 |
| African American | 42.68 |
| Others | 8.78 |

Table 2.1 Descriptive statistics stratified by COPD disease at Wave-1

| Variables | COPD Disease (Yes) | COPD Disease (No) | p-value |
|---|---|---|---|
| Wave 1 (n=168542) | Yes(n=21712) | No(n=146830) | |
| **Tobacco Use** | | | |
| Yes | 0 | 76730 | <0.001 |
| No | 21712 | 70100 | |
| **Wave 2 or Wave 3 Tobacco Use without COPD in Wave1** | COPD Disease either Wave 2 or Wave-3 (Yes) | COPD Disease either Wave 2 or Wave-3 (No) | |
| Yes | 3376 | 51553 | <0.001 |
| No | 6949 | 50596 | |
| **Covariates at Wave 1** | | | |
| **Comorbidities** | | | |
| **Diabetes Disease** | | | |
| Yes | 13522 | 56199 | <0.001 |
| No | 8190 | 90631 | |
| **COVID-19 Disease** | | | |
| Yes | 2686 | 17185 | 0.0046 |
| No | 19026 | 129645 | |
| **CAD Disease** | | | |
| Yes | 3302 | 5979 | <0.001 |
| No | 18410 | 140851 | |
| **Hypertension** | | | |
| Yes | 4143 | 6383 | <0.001 |
| No | 17569 | 140447 | |
| **Demographic** | | | |
| Age (years) | | | <0.001 |
| 15-25 | 77 | 23930 | |
| 26-35 | 593 | 38437 | |
| 36-45 | 2259 | 32629 | |
| 46-55 | 5967 | 25758 | |
| 56-65 | 10065 | 22025 | |
| 66-75 | 2751 | 4051 | |
| **Gender** | | | 0.821 |
| Male | 8237 | 58821 | |
| Female | 13475 | 88009 | |
| **Race** | | | <0.001 |
| White | 12819 | 68996 | |
| African American | 7076 | 64859 | |
| Others | 1817 | 12975 | |

14Table 2.2 Descriptive statistics stratified by COVID-19 disease at wave-1

| Variables | COVID-19 Disease (Yes) | COVID-19 Disease (No) | p-value |
|---|---|---|---|
| Wave 1 (n=168542) | Yes(n=19871) | No(n=148671) | |
| **Tobacco Use** | | | |
| Yes | 0 | 76730 | 0 |
| No | 19871 | 71941 | |
| **Wave 2 or Wave 3** | | | |
| **Tobacco Use among patients without COVID-19 in Wave1** | COVID-19 Disease either Wave 2 or Wave-3 (Yes) | COVID-19 Disease either Wave 2 or Wave-3 (No) | |
| Yes | 76502 | 228 | 0.73 |
| No | 71735 | 206 | |
| **Covariates at Wave 1** | | | |
| **Comorbidities** | | | |
| **Diabetes Disease** | | | |
| Yes | 8378 | 61343 | 0.016 |
| No | 11493 | 87328 | |
| **COPD Disease** | | | |
| Yes | 2686 | 19026 | 0.0046 |
| No | 17185 | 129645 | |
| **CAD Disease** | | | |
| Yes | 1232 | 8049 | 5.48 |
| No | 18639 | 140622 | |
| **Hypertension** | | | |
| Yes | 1676 | 8850 | 6.68 |
| No | 18195 | 139821 | |
| **Demographic** | | | |
| **Age (years)** | | | <0.001 |
| 15-25 | 3452 | 20555 | |
| 26-35 | 4447 | 34583 | |
| 36-45 | 3638 | 31250 | |
| 46-55 | 3236 | 28489 | |
| 56-65 | 3774 | 28316 | |
| 66-75 | 1324 | 5478 | |
| **Gender** | | | 0.821 |
| Male | 7891 | 59167 | |
| Female | 11980 | 89504 | |
| **Race** | | | <0.001 |
| White | 8561 | 73254 | |
| African American | 9399 | 62536 | |
| Others | 1911 | 12881 | |

Table 3 Cross-sectional associations with Chronic Obstructive Pulmonary Disease and COVID-19 at wave-1 (baseline)

| Variables | Cross-sectional associations with Chronic Obstructive Pulmonary Disease (COPD) at Wave 1 (baseline) | | Cross-sectional associations with COVID-19 at Wave 1 (baseline) | |
|---|---|---|---|---|
| | AOR (95% CI) | p-value | AOR (95% CI) | p-value |
| **Clinical Features** | | | | |
| COVID-19 | 0.7 (0.663, 0.739) | <0.001 | N/A | |
| COPD | N/A | | 0.79 (0.751, 0.833) | <0.001 |
| Hypertension | 1.75 (1.670, 1.846) | <0.001 | 1.07 (1.008, 1.143) | <0.05 |
| CAD | 1.18 (1.121, 1.247) | <0.001 | 0.87 (0.811, 0.931) | <0.001 |
| Total Number of Clinical Services | 1.05 (1.043, 1.047) | <0.001 | 1.04 (1.038, 1.042) | <0.001 |
| **Demographic Features** | | | | |
| Age (years) | | | | |
| 15-25 | Ref | | Ref | |
| 26-35 | 2.05 (1.605, 2.633) | <0.001 | 0.83 (0.773, 0.889) | <0.001 |
| 36-45 | 3.60 (2.776, 4.655) | <0.001 | 0.82 (0.726, 0.916) | 0.001 |
| 46-55 | 4.02 (3.001, 5.382) | <0.001 | 0.88 (0.739, 1.040) | 0.129 |
| 56-65 | 3.59 (2.588, 4.978) | <0.001 | 1.10 (0.886, 1.370) | 0.384 |
| 66-75 | 3.80 (2.625, 5.501) | <0.001 | 1.73 (1.322, 2.277) | <0.001 |
| Gender | | | | |
| Female | Ref | | Ref | |
| Male | 1.46 (1.412, 1.510) | <0.001 | 1.05 (1.014, 1.080) | 0.005 |
| Race | | | | |
| White | Ref | | Ref | |
| African American | 0.48 (0.465, 0.500) | <0.001 | 1.21 (1.169, 1.244) | <0.001 |
| Disability | | | | |
| No | Ref | | Ref | |
| Yes | 1.92 (1.857, 1.996) | <0.001 | 0.82 (0.788, 0.852) | <0.001 |
| Regions | | | | |
| New Orleans | Ref | | Ref | |
| Baton Rouge | 0.89 (0.840, 0.947) | <0.001 | 1.39 (1.319, 1.464) | <0.001 |
| Thibodeax | 0.92 (0.863, 0.976) | <0.05 | 1.02 (0.956, 1.081) | 0.599 |
| Lafayette | 1.05 (0.994, 1.104) | 0.085 | 1.40 (1.338, 1.476) | <0.001 |
| Lake Charles | 1.66 (1.553, 1.784) | <0.001 | 1.68 (1.576, 1.799) | <0.001 |
| Alexandria | 1.45 (1.358, 1.543) | <0.001 | 1.51 (1.416, 1.602) | <0.001 |
| Shreveport | 1.23 (1.163, 1.298) | <0.001 | 1.51 (1.438, 1.592) | <0.001 |
| Monroe | 1.30 (1.220, 1.380) | <0.001 | 1.89 (1.788, 1.992) | <0.001 |

Table 4 Longitudinal associations of Tobacco usage along with Clinical Features and Demographic variables with Chronic Obstructive Pulmonary Disease and COVID-19 Infection

| | Longitudinal Association between incident chronic obstructive pulmonary disease (at Wave 2 or 3) and tobacco user at Wave 1 excluding people who were diagnosed with COPD at Wave 1 | | Longitudinal Association between incident COVID-19 disease (at Wave 2 or 3) and tobacco user at Wave 1 excluding people who were infected with COVID-19 at Wave 1 | |
|---|---|---|---|---|
| **Variables** | AOR (95% CI) | p-value | AOR (95% CI) | p-value |
| Tobacco Usage | | | | |
| Yes | 1.12 (1.05, 1.18) | <0.001 | 1.66 (1.15, 2.38) | <0.01 |
| Clinical Features | | | | |
| COVID-19 | 0.84 (0.76, 0.92) | <0.001 | N/A | N/A |
| COPD | N/A | N/A | 1.04 (0.66, 1.62) | 0.877 |
| Hypertension | 1.58 (1.43, 1.73) | <0.001 | 1.11 (0.57, 2.15) | 0.763 |
| CAD | 1.10 (1.00, 1.22) | 0.059 | 1.13 (0.58, 2.19) | 0.725 |
| Total Number of Clinical Services | 1.02 (1.01, 1.03) | <0.05 | 1.03 (0.97, 1.10) | 0.295 |
| **Demographic Features** | | | | |
| Age (years) | | | | |
| 15-25 | Ref | | | |
| 26-35 | 1.96 (1.57, 2.46) | <0.001 | 1.23 (0.55, 2.79) | 0.612 |
| 36-45 | 2.87 (2.22, 3.73) | <0.001 | 2.27 (0.68, 7.52) | 0.181 |
| 46-55 | 3.44 (2.49, 4.76) | <0.001 | 3.33 (0.62, 18.07) | 0.163 |
| 56-65 | 2.94 (1.99, 4.33) | <0.001 | 4.90 (0.58, 41.26) | 0.144 |
| 66-75 | 3.42 (2.15, 5.42) | <0.001 | 16.11 (1.07, 243.72) | 0.045 |
| Gender | | | | |
| Female | Ref | | | |
| Male | 0.89 (0.85, 0.94) | <0.001 | 0.71 (0.53, 0.96) | <0.05 |
| Race | | | | |
| White | Ref | | | |
| African American | 0.58 (0.55, 0.62) | <0.001 | 1.07 (0.78, 1.48) | 0.681 |
| Disability | | | | |
| No | Ref | | | |
| Yes | 1.81 (1.72, 1.91) | <0.001 | 1.04 (0.74, 1.46) | 0.817 |
| Regions | | | | |
| New Orleans | Ref | | | |
| Baton Rouge | 1.02 (0.93, 1.12) | 0.687 | 0.58 (0.32, 1.04) | 0.07 |
| Thibodeax | 0.97 (0.88, 1.07) | 0.569 | 0.67 (0.35, 1.27) | 0.216 |
| Lafayette | 1.34 (1.23, 1.46) | <0.001 | 0.44 (0.25, 0.75) | <0.01 |
| Lake Charles | 2.10 (1.88, 2.33) | <0.001 | 0.86 (0.36, 2.04) | 0.733 |
| Alexandria | 1.92 (1.74, 2.12) | <0.001 | 0.57 (0.29, 1.13) | 0.105 |
| Shreveport | 1.52 (1.39, 1.66) | <0.001 | 0.64 (0.35, 1.19) | 0.161 |
| Monroe | 1.48 (1.34, 1.63) | <0.001 | 0.55 (0.29, 1.04) | 0.065 |
| Mandeville | 1.30 (1.19, 1.42) | <0.001 | 0.54 (0.30, 0.96) | <0.05 |

For COVID-19, important correlates at Wave-1 included hypertension (AOR=1.07, 95% CI=1.008–1.143) and total clinical visits (AOR=1.04, 95% CI=1.038–1.042). Individuals aged 66–75 (AOR=1.73, 95% CI=1.32–2.28) and those living in Monroe (AOR=1.89, 95% CI=1.79–1.99) were also more likely to have COVID-19 at baseline.

Two longitudinal logistic regression models were constructed as shown in Table 4 to capture new (incident) cases of COPD (left column) and COVID-19 (right column) through Waves 2 and 3. Among enrollees who did not have COPD at Wave-1, tobacco use at baseline was linked to higher odds of developing COPD later (AOR=1.12, 95% CI=1.05–1.18, p<0.001). Hypertension (AOR=1.58, 95% CI=1.43–1.73), CAD (AOR=1.10, 95% CI=1.00–1.22), and total clinical services (AOR=1.02, 95% CI=1.01–1.03) also showed positive associations with COPD incidence. Additionally, disability status (AOR=1.81, 95% CI=1.72–1.91), the 46–55 age group (AOR=3.44, 95% CI=2.49–4.76), and residing in Lake Charles (AOR=2.10, 95% CI=1.88–2.33) further elevated COPD risk.

For the COVID-19 model, baseline tobacco use significantly increased the likelihood of subsequent infection (AOR=1.66, 95% CI=1.15–2.38, p<0.01). Among age cohorts, only 66–75 showed a significant impact on new COVID-19 diagnoses (AOR=16.11, 95% CI=1.07–243.72), although this estimate had a wide confidence interval. The Lafayette region exhibited lower odds of incident infection compared to New Orleans (AOR=0.44, 95% CI=0.25–0.75), and men had reduced odds relative to women (AOR=0.71, p<0.05). In contrast to the COPD model, neither disability status nor the total number of clinical services predicted COVID-19 incidence over the follow-up period.

Receiver operating characteristic (ROC) curves, shown in Figure 1(C), evaluated the predictive accuracy of each longitudinal model. The COPD model achieved an AUROC of 0.80, denoting strong discriminatory power, while the COVID-19 model reached an AUROC of 0.65, reflecting modest predictive capability. Collectively, these findings underscore the multifaceted role of tobacco use, comorbidities, and demographic factors in shaping COPD and COVID-19 risk among Louisiana Medicaid enrollees.

## 4. Discussions

This research is among the first to focus specifically on Louisiana Medicaid enrollees, illuminating how tobacco consumption, comorbid conditions, and social determinants of health jointly influence COPD and COVID-19 outcomes.

In our baseline (Wave 1) cross-sectional analysis, hypertension (AOR = 1.75, 95% CI = 1.67–1.85) and increased clinical service utilization (AOR = 1.05, 95% CI = 1.04- 1.05, $p < 0.001$) were both associated with greater COPD risk. These findings align with NHANES data showing 96.4% of adults with COPD have at least one other chronic condition- most notably hypertension (60.4%)- that intensifies COPD severity [26]. The link between rising COPD prevalence and higher clinical visits (Table 3) also resonates with evidence that Medicaid beneficiaries who have COPD tend to rely heavily on medical services [27]. Additionally, factors such as age 46–55, male sex, disability status, and residence in the Lake Charles region emerged as significant predictors of COPD, consistent with existing studies identifying similar risk profiles [28,29,30].

For COVID-19 at Wave 1, patients with hypertension (AOR = 1.07, 95% CI = 1.008–1.143) and those seeking more clinical services (AOR = 1.04, 95% CI = 1.038–1.042) exhibited slightly higher odds of infection. These results mirror a previous study [31] in which hospitalized COVID-19 patients—especially those with heightened anxiety—had strong links to hypertension. Another investigation [32] demonstrated disparities in COVID-19 incidence across multiple demographic groups, echoing the patterns observed in our analysis.

Turning to longitudinal trends, the first model reveals that baseline tobacco use increased the odds (AOR = 1.12, 95% CI = 1.05–1.18) of developing COPD in the following waves, consistent with broader research on smoking-related respiratory harm [33–36]. Although this effect size is somewhat lower than estimates from Bhatta and Glantz [37] for e-cigarettes (AOR = 1.29) and combustible tobacco (AOR = 2.56), our data also highlight additional risk factors—CAD, hypertension, and extensive medical utilization (Table 4)—that potentiate COPD onset. Demographic factors such as age 46–55 (AOR = 3.44, 95% CI = 2.49–4.76) similarly mirror Westney et al. [38], who found higher COPD-related clinical visits among adults aged 45–64. Meanwhile, White enrollees showed an increased COPD likelihood relative to African Americans (AOR = 0.58, 95% CI = 0.55–0.62), tracking with national trends [38]. Disability status (AOR = 1.81, 95% CI = 1.72–1.91) also emerged as a considerable risk factor [39], and women in this population were more prone than men to COPD (AOR = 0.89, 95% CI = 0.85–0.94)—an observation consistent with previous work indicating females have an 18% greater likelihood of hospitalization for COPD and asthma [40]. Overall, these socio-demographic patterns align closely with broader national data [34,36,41].

In the second longitudinal model, tobacco use again stood out as a principal risk factor—this time for COVID-19 infection (AOR = 1.66, 95% CI = 1.15–2.38, p < 0.01). This odds ratio is higher than that for COPD among these enrollees but still lower than certain youth-oriented estimates showing markedly elevated COVID-19 risk for e-cigarette and dual users (AOR = 5.05 and AOR = 6.97, respectively) [42], and higher than the 0.64 (95% CI = 0.49–0.84) identified in a sample of navy aircraft personnel [43]. It also surpasses reported AORs for ever smokers (1.26), current smokers (1.23), and former smokers (1.28) in another study examining COVID-19 severity [44]. Regardless, the link between smoking and susceptibility to COVID-19 concurs with evidence that tobacco exposure heightens ACE2 expression in the respiratory epithelium [7,45–50].

A major strength of our investigation lies in leveraging extensive Louisiana Medicaid claims data, which allowed for a robust examination of how tobacco use, chronic conditions, and socio-demographic contexts shape disease pathways over time. Although prior research has examined COPD and COVID-19 separately within certain subpopulations, our longitudinal approach integrates these dimensions specifically for Medicaid enrollees in Louisiana, generating important insights into how public health interventions might reduce the compounded burden of respiratory illness in this vulnerable group.

## 5. Limitations

Although this study provides valuable insights, several limitations should be acknowledged. First, disease identification relied on aggregated ICD-10 codes, which did not distinguish between combustible and non-combustible tobacco use—potentially obscuring important nuances. Second, we used the raw dataset without employing balancing techniques for the target diseases, making the distribution of COPD and COVID-19 inherently imbalanced and possibly introducing bias. Finally, our analysis was restricted to Medicaid enrollees in Louisiana, limiting the generalizability of these findings to broader populations or to Medicaid programs in other states.

6. Conclusion

In conclusion, baseline tobacco use emerged as a robust and independent predictor for both COPD and COVID-19 in the Louisiana Medicaid cohort. Hypertension consistently demonstrated a significant relationship with COPD across cross-sectional and longitudinal analyses, while being over age 45 elevated risk for both conditions. For COPD, additional factors such as female gender, White race, higher healthcare utilization, and residence in specific regions were strongly linked to increased disease incidence. In contrast, demographic variables—particularly gender and the Lafayette region—stood out in the COVID-19 model. The COPD model's AUROC of 0.80 indicates strong predictive accuracy, whereas the COVID-19 model's AUROC of 0.65 signals a moderate but meaningful level of discrimination. These findings underscore the need for tailored interventions, especially tobacco-cessation efforts, within the Louisiana Medicaid population. By leveraging these insights, clinicians and policymakers can enhance strategies to mitigate COPD and COVID-19 risks and ultimately improve respiratory outcomes among Medicaid enrollees.

7. Acknowledgements

This study was supported in part by the Public University Partnership Program at the Louisiana Department of Health, Bureau of Health Services Financing.

**Disclaimer**: This content is the sole responsibility of the authors and does not necessarily represent the official views of the Louisiana Department of Health.

8. **Declaration of Interest Statement**

The authors declare that they have no known competing financial interests or personal relationships that could have appeared to influence the work reported in this paper.

9. **References**

1. Rosenberg SR, Kalhan R, Mannino DM. Epidemiology of Chronic Obstructive Pulmonary Disease: Prevalence, Morbidity, Mortality, and Risk Factors. *Semin Respir Crit Care Med*. 2015;36(4):457-469. doi:10.1055/S-0035-1555607

2. Liu Y, Carlson SA, Watson KB, Xu F, Greenlund KJ. Trends in the Prevalence of Chronic Obstructive Pulmonary Disease Among Adults Aged ≥18 Years — United States, 2011–2021. *MMWR Morb Mortal Wkly Rep*. 2023;72(46):1250-1256. doi:10.15585/MMWR.MM7246A1

3. COPD Trends Brief - Mortality | American Lung Association. Accessed April 29, 2024. https://www.lung.org/research/trends-in-lung-disease/copd-trends-brief/copd-mortality

4. CDC Museum COVID-19 Timeline | David J. Sencer CDC Museum | CDC. Accessed April 29, 2024. https://www.cdc.gov/museum/timeline/covid19.html

5. CDC COVID Data Tracker: Maps by Geographic Area. Accessed April 30, 2024. https://covid.cdc.gov/covid-data-tracker/#maps_positivity-week

6. Münzel T, Hahad O, Kuntic M, Keaney JF, Deanfield JE, Daiber A. Effects of tobacco cigarettes, e-cigarettes, and waterpipe smoking on endothelial function and clinical outcomes. *Eur Heart J*. 2020;41(41):4057-4070. doi:10.1093/EURHEARTJ/EHAA460

7. Leung JM, Yang CX, Tam A, et al. ACE-2 expression in the small airway epithelia of smokers and COPD patients: implications for COVID-19. *European Respiratory Journal*. 2020;55(5). doi:10.1183/13993003.00688-2020

8. Behr J, Nowak D. Tobacco smoke and respiratory disease. Published online 2002. Accessed April 29, 2024. https://www.researchgate.net/publication/251883915

9. Song Q, Chen P, Liu XM. The role of cigarette smoke-induced pulmonary vascular endothelial cell apoptosis in COPD. *Respiratory Research 2021 22:1*. 2021;22(1):1-15. doi:10.1186/S12931-021-01630-1

10. Moretto N, Volpi G, Pastore F, Facchinetti F. Acrolein effects in pulmonary cells: relevance to chronic obstructive pulmonary disease. *Ann N Y Acad Sci*. 2012;1259(1):39-46. doi:10.1111/J.1749-6632.2012.06531.X

11. Gaffney AW, Hawks L, White AC, et al. Health Care Disparities Across the Urban-Rural Divide: A National Study of Individuals with COPD. *The Journal of Rural Health*. 2022;38(1):207-216. doi:10.1111/JRH.12525

12. Cornelius ME, Wang TW, Jamal A, et al. State-Specific Prevalence of Adult Tobacco Product Use and Cigarette Smoking Cessation Behaviors, United States, 2018–2019. *Prev Chronic Dis*. 2023;20. doi:10.5888/PCD20.230132

13. Babb S, Malarcher A, Schauer G, Asman K, Jamal A. Quitting Smoking Among Adults — United States, 2000–2015. *MMWR Morb Mortal Wkly Rep*. 2017;65(52):1457-1464. doi:10.15585/mmwr.mm6552a1

14. Zhu SH, Anderson CM, Wong S, Kohatsu ND. The Growing Proportion of Smokers in Medicaid and Implications for Public Policy. *Am J Prev Med*. 2018;55(6):S130-S137. doi:10.1016/J.AMEPRE.2018.07.017

15. Armour BS, Finkelstein EA, Fiebelkorn IC. Peer Reviewed: State-Level Medicaid Expenditures Attributable to Smoking. *Prev Chronic Dis*. 2009;6(3). Accessed April 30, 2024. /pmc/articles/PMC2722402/

16. Chapel JM, Ritchey MD, Zhang D, Wang G. Prevalence and Medical Costs of Chronic Diseases Among Adult Medicaid Beneficiaries. *Am J Prev Med*. 2017;53(6):S143-S154. doi:10.1016/J.AMEPRE.2017.07.019

17. Shao Q, Polavarapu M, Small L, Singh S, Nguyen Q, Shao K. A longitudinal mixed effects model for assessing mortality trends during vaccine rollout. *Healthcare Analytics*. 2024;6:100347. doi:10.1016/J.HEALTH.2024.100347

18. Aanegola R, Nakamura Sakai S, Kumar N. Longitudinal analysis of the determinants of life expectancy and healthy life expectancy: A causal approach. *Healthcare Analytics*. 2022;2:100028. doi:10.1016/J.HEALTH.2022.100028

19. Gonzalez-Canas C, Pujol TA, Griffin P, Hass Z. A multilevel logistic regression model for identifying the relevance of environmental risk factors on Gestational Diabetes Mellitus. *Healthcare Analytics*. 2023;3:100152. doi:10.1016/J.HEALTH.2023.100152

20. Mori M, Flores RG, Kamimura H, Yamaura K, Nonaka H. An analytical investigation of body parts more susceptible to aging and composition changes using statistical hypothesis testing. *Healthcare Analytics*. 2024;5:100284. doi:10.1016/J.HEALTH.2023.100284

21. Pincombe A, Mittinty MN, Karnon J. A scoping review of the statistical methods and risk-adjustment approaches used to compare cardiovascular disease services using Australian

health system data. *Healthcare Analytics*. 2023;4:100250. doi:10.1016/J.HEALTH.2023.100250

22. Archived ICD-10-CM Diagnosis Code Policy Charts | La Dept. of Health. Accessed May 1, 2024. https://ldh.la.gov/page/archived-icd10cm-diagnosis-code-policy-charts

23. Fávero LP, Belfiore P, de Freitas Souza R. Binary and multinomial logistic regression models. *Data Science, Analytics and Machine Learning with R*. Published online January 1, 2023:259-283. doi:10.1016/B978-0-12-824271-1.00008-1

24. Regression with Discrete Dependent Variable - statsmodels 0.14.1. Accessed May 1, 2024. https://www.statsmodels.org/stable/discretemod.html

25. Hoetker G. The use of logit and probit models in strategic management research: Critical issues. *Strategic Management Journal*. 2007;28(4):331-343. doi:10.1002/SMJ.582

26. Schnell K, Weiss CO, Lee T, et al. The prevalence of clinically-relevant comorbid conditions in patients with physician-diagnosed COPD: A cross-sectional study using data from NHANES 1999-2008. *BMC Pulm Med*. 2012;12(1):1-9. doi:10.1186/1471-2466-12-26/TABLES/3

27. Lin PJ, Shaya FT, Scharf SM. Economic implications of comorbid conditions among Medicaid beneficiaries with COPD. *Respir Med*. 2010;104(5):697-704. doi:10.1016/J.RMED.2009.11.009

28. Woldeamanuel GG, Mingude AB, Geta TG. Prevalence of chronic obstructive pulmonary disease (COPD) and its associated factors among adults in Abeshge District, Ethiopia: A cross sectional study. *BMC Pulm Med*. 2019;19(1):1-9. doi:10.1186/S12890-019-0946-Z/TABLES/5

29. Avcı S, Üniversitesi U. STATISTICAL EVALUATION OF COPD PATIENTS WITH RESPECT TO GENDER: A CROSS-SECTIONAL STUDY. *Baqai J Health Sci*. 21(2). Accessed May 4, 2024. https://www.researchgate.net/publication/330715746

30. Yohannes AM. Disability in patients with COPD. *Chest*. 2014;145(2):200-202. doi:10.1378/chest.13-1703

31. Sensoy B, Gunes A, Ari S. Anxiety and depression levels in Covid-19 disease and their relation to hypertension. *Clin Exp Hypertens*. 2021;43(3):237-241. doi:10.1080/10641963.2020.1847132

32. Theodore DA, Branche AR, Zhang L, et al. Clinical and Demographic Factors Associated With COVID-19, Severe COVID-19, and SARS-CoV-2 Infection in Adults: A Secondary Cross-Protocol Analysis of 4 Randomized Clinical Trials. *JAMA Netw Open*. 2023;6(7):e2323349-e2323349. doi:10.1001/JAMANETWORKOPEN.2023.23349

33. Sargent JD, Halenar MJ, Edwards KC, et al. Tobacco Use and Respiratory Symptoms Among Adults: Findings From the Longitudinal Population Assessment of Tobacco and Health (PATH) Study 2014–2016. *Nicotine & Tobacco Research*. 2022;24(10):1607-1618. doi:10.1093/NTR/NTAC080

34. Paulin LM, Halenar MJ, Edwards KC, et al. Association of tobacco product use with chronic obstructive pulmonary disease (COPD) prevalence and incidence in Waves 1 through 5 (2013–2019) of the Population Assessment of Tobacco and Health (PATH) Study. *Respir Res*. 2022;23(1):1-13. doi:10.1186/S12931-022-02197-1/TABLES/4

35. Cook S, Buskiewicz J, Levy DT, Meza R, Fleischer NL. Association between cigar use, with and without cigarettes, and incident diagnosed COPD: a longitudinal cohort study. *Respir Res*. 2024;25(1):1-9. doi:10.1186/S12931-023-02649-2/TABLES/4

36. Cunningham TJ, Ford ES, Rolle I V., Wheaton AG, Croft JB. Associations of Self-Reported Cigarette Smoking with Chronic Obstructive Pulmonary Disease and Co-Morbid Chronic Conditions in the United States. *COPD: Journal of Chronic Obstructive Pulmonary Disease*. 2015;12(3):281-291. doi:10.3109/15412555.2014.949001

37. Bhatta DN, Glantz SA. Association of E-Cigarette Use With Respiratory Disease Among Adults: A Longitudinal Analysis. *Am J Prev Med*. 2020;58(2):182-190. doi:10.1016/J.AMEPRE.2019.07.028

38. Westney G, Foreman MG, Xu J, King MH, Flenaugh E, Rust G. Peer Reviewed: Impact of Comorbidities Among Medicaid Enrollees With Chronic Obstructive Pulmonary Disease, United States, 2009. *Prev Chronic Dis*. 2017;14(4):160333. doi:10.5888/PCD14.160333

39. Thornton Snider J, Romley JA, Wong KS, Zhang J, Eber M, Goldman DP. The Disability Burden of COPD. *COPD: Journal of Chronic Obstructive Pulmonary Disease*. 2012;9(5):513-521. doi:10.3109/15412555.2012.696159

40. Shaya FT, Dongyi D, Akazawa MO, et al. Burden of Concomitant Asthma and COPD in a Medicaid Population*. *Chest*. 2008;134(1):14-19. doi:10.1378/CHEST.07-2317

41. Tilert T, Paulose-Ram R, Howard D, Butler J, Lee S, Wang MQ. Prevalence and factors associated with self-reported chronic obstructive pulmonary disease among adults aged 40-79: the National Health and Nutrition Examination Survey (NHANES) 2007–2012. *EC Pulmonol Respir Med*. 2018;7(9):650. Accessed April 29, 2024. /pmc/articles/PMC6169793/

42. Gaiha SM, Cheng J, Halpern-Felsher B. Association Between Youth Smoking, Electronic Cigarette Use, and COVID-19. *J Adolesc Health*. 2020;67(4):519-523. doi:10.1016/J.JADOHEALTH.2020.07.002

43. Paleiron N, Mayet A, Marbac V, et al. Impact of Tobacco Smoking on the Risk of COVID-19: A Large Scale Retrospective Cohort Study. *Nicotine Tob Res*. 2021;23(8):1398-1404. doi:10.1093/NTR/NTAB004

44. Mahabee-Gittens EM, Mendy A, Merianos AL. Assessment of severe COVID-19 outcomes using measures of smoking status and smoking intensity. *Int J Environ Res Public Health*. 2021;18(17). doi:10.3390/IJERPH18178939/S1

45. Li S, Han J, Zhang A, et al. Exploring the Demographics and Clinical Characteristics Related to the Expression of Angiotensin-Converting Enzyme 2, a Receptor of SARS-CoV-2. *Front Med (Lausanne)*. 2020;7:530. doi:10.3389/FMED.2020.00530

46. Smith JC, Sausville EL, Girish V, et al. Cigarette Smoke Exposure and Inflammatory Signaling Increase the Expression of the SARS-CoV-2 Receptor ACE2 in the Respiratory Tract. *Dev Cell*. 2020;53:514-529.e3. doi:10.1016/j.devcel.2020.05.012

47. Brake SJ, Barnsley K, Lu W, McAlinden KD, Eapen MS, Sohal SS. Smoking Upregulates Angiotensin-Converting Enzyme-2 Receptor: A Potential Adhesion Site for Novel Coronavirus SARS-CoV-2 (Covid-19). *Journal of Clinical Medicine 2020, Vol 9, Page 841*. 2020;9(3):841. doi:10.3390/JCM9030841

48. Choi JY, Lee HK, Park JH, et al. Altered COVID-19 receptor ACE2 expression in a higher risk group for cerebrovascular disease and ischemic stroke. *Biochem Biophys Res Commun*. 2020;528(3):413-419. doi:10.1016/J.BBRC.2020.05.203

49. Lee AC, Chakladar J, Li WT, et al. Tobacco, but Not Nicotine and Flavor-Less Electronic Cigarettes, Induces ACE2 and Immune Dysregulation. *International Journal of Molecular Sciences 2020, Vol 21, Page 5513*. 2020;21(15):5513. doi:10.3390/IJMS21155513

50. Saheb Sharif-Askari N, Saheb Sharif-Askari F, Alabed M, et al. Airways Expression of SARS-CoV-2 Receptor, ACE2, and TMPRSS2 Is Lower in Children Than Adults and Increases with Smoking and COPD. *Mol Ther Methods Clin Dev*. 2020;18:1-6. doi:10.1016/j.omtm.2020.05.013